\documentclass[amssymb,pra,twocolumn,notitlepage,superscriptaddress,longbibliography,floatfix]{revtex4-2}
\usepackage[T1]{fontenc}
\usepackage{amsmath}
\usepackage{amssymb}
\usepackage{graphicx}
\usepackage{color}
\usepackage[unicode=true,pdfusetitle,
 bookmarks=true,bookmarksnumbered=false,bookmarksopen=false,
 breaklinks=false,pdfborder={0 0 0},backref=false,colorlinks=true]
 {hyperref}

\begin{document}

\title{Interference Engineering for Quantum Imaginary-Time Evolution through Multiple Energy Shifts}

\author{Hong-Jian Tang}
\affiliation{Key Laboratory of Atomic and Subatomic Structure and Quantum Control (Ministry of Education),  Guangdong Basic Research Center of Excellence for Structure and Fundamental Interactions of Matter, and  School of Physics, South China Normal University, Guangzhou 510006, China}

\author{Dan-Bo Zhang}
\email{dbzhang@m.scnu.edu.cn}
\affiliation{Key Laboratory of Atomic and Subatomic Structure and Quantum Control (Ministry of Education),  Guangdong Basic Research Center of Excellence for Structure and Fundamental Interactions of Matter, and  School of Physics, South China Normal University, Guangzhou 510006, China}
\affiliation{Guangdong Provincial Key Laboratory of Quantum Engineering and Quantum Materials,  Guangdong-Hong Kong Joint Laboratory of Quantum Matter, and Frontier Research Institute for Physics,\\  South China Normal University, Guangzhou 510006, China}

\date{\today}

\begin{abstract}
Energy shifting is usually trivial in imaginary-time evolution because it changes only the normalization of the evolved state. On a quantum computer, however, imaginary-time evolution can be implemented as a coherent or sampled superposition of real-time evolutions, in which energy shifts generate relative phases that can interfere. Here we introduce multi-shift quantum imaginary-time evolution (MS-QITE), which uses a distribution of energy shifts to engineer this interference and optimize different implementations. In a Monte Carlo realization, multi-shift reshapes the normalized sampling distribution and concentrates it within a shorter real-time window, thereby reducing the typical Hamiltonian-evolution time and improving the stability of ground-state-energy estimation. In a continuous-variable-assisted realization, it enables projection onto a state supported over a finite quadrature interval, substantially reducing the required squeezing over an intermediate temperature range while retaining accurate thermal-state preparation. Numerical results for transverse-field Ising models demonstrate that energy shifts provide an interference-based degree of freedom for optimizing quantum imaginary-time evolution.
\end{abstract}

\maketitle

\section{Introduction}

Imaginary-time evolution (ITE) is a fundamental tool in quantum many-body physics and quantum chemistry~\cite{LEHTOVAARA2007148,McArdle2018VariationalAQ,Motta2019DeterminingEA}. By exponentially suppressing excited-state components, it provides a direct route to ground-state preparation, thermal-state preparation, and finite-temperature simulation~\cite{Poulin2008PreparingGS,Motta2019DeterminingEA,PRXQuantum.2.010317,Temme2009QuantumMS,Poulin2009SamplingFT,Wu2018VariationalTQ}. In classical computation, ITE underlies or is closely connected to widely used approaches including quantum Monte Carlo and tensor-network methods~\cite{Sorella2000GeneralizedLA,Lester1990QuantumMC,PhysRevLett.82.4745,Verstraete01032008}.

For a Hamiltonian $H$ and imaginary time $\tau$, implementing ITE on a quantum computer is less direct because the operator $e^{-H\tau}$ is nonunitary and therefore cannot be realized as an ordinary gate-based evolution~\cite{McArdle2018VariationalAQ}. A broad range of strategies has therefore been developed. Variational formulations include variational QITE and related variational simulation principles~\cite{McArdle2018VariationalAQ,Yuan2018TheoryOV}. Local-unitary and circuit-based QITE approaches include the original QITE construction, practical QITE--Lanczos implementations, compressed-circuit evolution, double-bracket flows, Riemannian-gradient-flow constructions, and multi-copy unitary protocols~\cite{Motta2019DeterminingEA,YeterAydeniz2020PracticalQITE,PRXQuantum.2.010342,rw81-k8vk,McMahon2026Riemannian,Schwartzman2026Multicopy}. Deterministic or probabilistic realizations of the nonunitary map provide another route~\cite{PhysRevLett.131.110602,Liu2021}, while Gibbs-state and finite-temperature preparation have been pursued through variational and sampling-based constructions~\cite{PhysRevApplied.16.054035,Sagastizabal2020VariationalPO,Temme2009QuantumMS,Poulin2009SamplingFT,Wu2018VariationalTQ}. Other approaches include linear-combination-of-unitaries methods~\cite{Childs2012HamiltonianSU,Chowdhury2016QuantumAF,PhysRevLett.131.150603} and Monte Carlo, algorithmic-cooling, and real-time-filtering methods~\cite{Huo2021ErrorresilientMC,zeng2022universalquantumalgorithmiccooling,PhysRevA.109.052416,Zhang2024measurement,PhysRevA.106.032420}. Moreover, imaginary-time evolution operator can also be constructed by one assisted continuous variable with post-selection ~\cite{Zhang2020ContinuousVariableAT, Bell2025SpectralOracles}

Monte Carlo methods and continuous-variable assisted methods provide two representative routes for implementing ITE through real-time quantum evolution. However, both methods still suffer from practical limitations. In Monte Carlo approaches, the imaginary-time evolution is expressed as a weighted superposition of real-time evolutions, where the sampling distribution possesses a long-time tail. Consequently, long real-time evolutions are frequently sampled, resulting in increased circuit depth and more severe noise accumulation on quantum hardware~\cite{Huo2021ErrorresilientMC}. In continuous-variable-assisted methods, imaginary-time evolution is realized through the postselection of an auxiliary continuous-variable state. The choice of the projection state directly affects the success probability of the algorithm~\cite{Zhang2020ContinuousVariableAT}. Designing more suitable projection states to improve the success probability without changing the imaginary-time evolution is therefore of practical interest.

A possible way to improve these algorithms comes from the idea of energy shifting. Energy shifts have been widely used in classical numerical algorithms, such as shift-and-invert methods for eigenvalue problems, where they modify spectral properties to improve computational performance~\cite{Sorensen2002NumericalMF,Schreiber2007AJL}. For conventional imaginary-time evolution, however, adding a constant energy shift only introduces an overall normalization factor and therefore does not change the final normalized state. Consequently, energy shifts have rarely been explored as a tool for improving imaginary-time evolution itself. On quantum computers, the situation is fundamentally different. Since many quantum implementations express imaginary-time evolution as a superposition of real-time evolutions~\cite{Kosugi2021ImaginarytimeEU}, an energy shift appears as an additional phase factor in the real-time representation rather than merely an overall normalization. This observation provides additional freedom for algorithm design and opens up new possibilities for optimizing quantum implementations of imaginary-time evolution.

Motivated by this observation, we propose multi-shift quantum imaginary-time evolution (MS-QITE). Instead of applying a single constant energy shift, we introduce a distribution of energy shifts characterized by a function $f(q)$. Although the normalized imaginary-time evolution remains unchanged, the resulting phase factor in the real-time representation can be tailored to optimize different quantum implementations. We demonstrate the application of the proposed approach to both Monte Carlo and continuous-variable-assisted imaginary-time evolution. For Monte Carlo methods, an appropriate multi-shift function reshapes the sampling distribution and shortens the effective real-time evolution, thereby reducing circuit depth and noise accumulation. For continuous-variable-assisted methods, multi-shift is realized through the choice of the projection state, thereby reducing resource consumption while maintaining high preparation accuracy. Numerical simulations of ground-state preparation and thermal-state preparation verify the effectiveness of the proposed approach. These results show that multi-shift provides a flexible algorithmic perspective for improving quantum imaginary-time evolution and may be applicable to a broader class of quantum algorithms.

The remainder of this paper is organized as follows. Sec.~\ref{s2} introduces the multi-shift formulation. Sec.~\ref{s3} develops its Monte Carlo implementation and demonstrates the reduction of the required real-time evolution. Sec.~\ref{s4} presents the continuous-variable-assisted implementation and its application to thermal-state preparation. Sec.~\ref{s5} concludes the paper.

\section{Multi-Shift Formulation}
\label{s2}

Let $H$ be a Hermitian Hamiltonian with nonnegative spectrum, $\tau>0$ the imaginary time, and $|\psi(0)\rangle$ a normalized initial state. If the physical Hamiltonian is not nonnegative, a constant multiple of the identity can first be added; this does not change the normalized imaginary-time-evolved state. The normalized evolution is
\begin{equation}\label{eq:ite}
|\psi(\tau)\rangle=
\frac{e^{-H\tau}|\psi(0)\rangle}
{\sqrt{\langle\psi(0)|e^{-2H\tau}|\psi(0)\rangle}}.
\end{equation}

For a single real energy shift $b$, replacing $H$ by $H+bI$, where $I$ is the identity operator, multiplies the unnormalized state by $e^{-b\tau}$ and therefore leaves Eq.~\eqref{eq:ite} unchanged after normalization. More generally, let $f(q)\geq 0$ be a normalized distribution of energy shifts $q$ on a support $\mathcal S$, with $\int_{\mathcal S}dq\,f(q)=1$. Provided that $H+qI$ has nonnegative spectrum for every $q\in\mathcal S$, one has
\begin{equation}\label{eq:multishift}
\begin{aligned}
\int_{\mathcal S}dq\,f(q)e^{-(H+qI)\tau}
&=C_f(\tau)e^{-H\tau},\\
C_f(\tau)&\equiv\int_{\mathcal S}dq\,f(q)e^{-q\tau}>0.
\end{aligned}
\end{equation}
Thus, the shift distribution changes only an overall scalar in imaginary time, while leaving the normalized state invariant.

For a nonnegative Hamiltonian, the scalar identity
$e^{-E\tau}=\int_{-\infty}^{\infty}dt\,g_\tau(t)e^{-iEt}$ for $E\geq0$ extends spectrally to
\begin{equation}\label{eq:cauchy}
e^{-H\tau}=
\int_{-\infty}^{\infty}dt\,g_\tau(t)e^{-iHt},
\qquad
g_\tau(t)\equiv \frac{\tau}{\pi(\tau^2+t^2)}.
\end{equation}
Here $g_\tau(t)$ is the normalized Cauchy density. Applying Eq.~\eqref{eq:cauchy} to each shifted Hamiltonian gives
\begin{equation}\label{eq:multishift-rt}
\int_{\mathcal S}dq\,f(q)e^{-(H+qI)\tau}
=
\int_{-\infty}^{\infty}dt\,g_\tau(t)W(t)e^{-iHt},
\end{equation}
where
\[
W(t)\equiv\int_{\mathcal S}dq\,f(q)e^{-iqt}
\]
is the characteristic function of the shift distribution. Equations~\eqref{eq:multishift} and \eqref{eq:multishift-rt} show the central point of MS-QITE: the same normalized imaginary-time evolution can be represented by different real-time interference kernels $g_\tau(t)W(t)$.

Before presenting numerical results, we emphasize that this freedom is not tied to a specific implementation. In Sec.~\ref{s3}, it modifies the sampling kernel of a quantum--classical Monte Carlo realization. In Sec.~\ref{s4}, the same function is encoded in the projection wavefunction of an auxiliary continuous-variable mode. These two examples illustrate complementary uses of the same interference degree of freedom.

\section{Multi-Shift Monte Carlo Implementation}
\label{s3}

Using Monte Carlo sampling, expectation values after imaginary-time evolution can be evaluated through a quantum--classical hybrid approach. For a Hermitian observable $O$ and an initial state $|\psi_0\rangle$, Eq.~\eqref{eq:cauchy} gives
\begin{equation}\label{eq:mc-expectation}
\langle O\rangle_\tau=
\frac{
\iint_{-\infty}^{\infty}dt\,dt'\,
g_\tau(t)g_\tau(t')
\langle\psi_0|e^{iHt'}Oe^{-iHt}|\psi_0\rangle
}{
\iint_{-\infty}^{\infty}dt\,dt'\,
g_\tau(t)g_\tau(t')
\langle\psi_0|e^{iHt'}e^{-iHt}|\psi_0\rangle
}.
\end{equation}
A finite cutoff $t_m$ is implemented by sampling from the corresponding distribution conditioned on $|t|\leq t_m$; its normalization cancels between the numerator and denominator, while the omitted tail produces a systematic cutoff error that vanishes as $t_m$ increases.

A practical estimator is obtained as follows:
\begin{enumerate}
\item \textit{On a classical computer, draw $N_s$ independent pairs $(t_l,t_l')$ from the chosen (possibly truncated) sampling distribution.}
\item \textit{On a quantum computer, prepare the initial state $|\psi_0\rangle$.}
\item \textit{For each pair, estimate
$\mu_l(O)\equiv\langle\psi_0|e^{iHt_l'}Oe^{-iHt_l}|\psi_0\rangle$
and
$\mu_l(I)\equiv\langle\psi_0|e^{iHt_l'}e^{-iHt_l}|\psi_0\rangle$.
For a unitary or Pauli observable these complex amplitudes can be obtained from Hadamard tests (real and imaginary parts separately); a general observable is first decomposed into a linear combination of unitary or Pauli terms.}
\item \textit{Form the ratio}
\begin{equation}\label{eq:mc-estimator}
\langle O\rangle_\tau
\approx
\frac{\displaystyle\sum_{l=1}^{N_s}\mu_l(O)}
{\displaystyle\sum_{l=1}^{N_s}\mu_l(I)}.
\end{equation}
\end{enumerate}

\begin{figure}[t]
\centering
\includegraphics[width=\columnwidth]{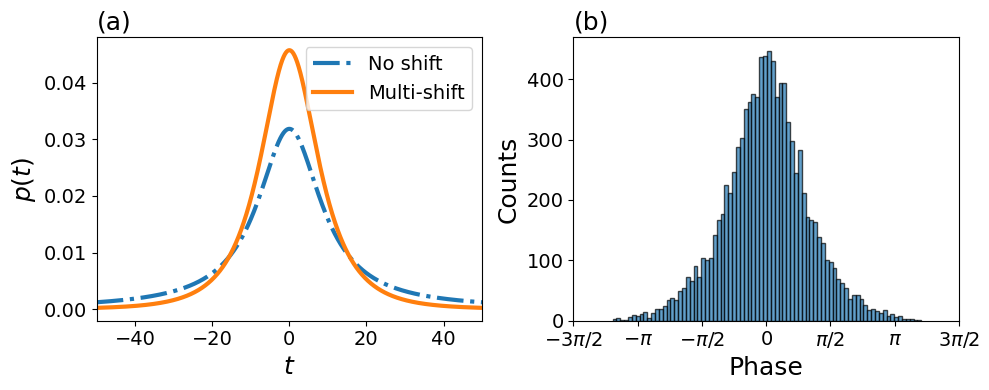}
\caption{(a) Normalized sampling density without and with multi-shift. (b) Histogram of the phase $\theta(t)$ for $10^4$ independent samples drawn from $p_W(t)$. The parameters are $a=30$, $k=3$, and $\tau=10$.}
\label{f1}
\end{figure}

With multi-shift, the modulus of $W(t)$ is absorbed into the sampling distribution. Because $g_\tau(t)|W(t)|$ is not normalized in general, define
\begin{equation}\label{eq:shift-sampling}
p_W(t)\equiv
\frac{g_\tau(t)|W(t)|}{Z_W},
\qquad
Z_W\equiv\int_{-\infty}^{\infty}dt\,g_\tau(t)|W(t)|.
\end{equation}
For $W(t)\neq0$, write $W(t)/|W(t)|=e^{i\theta(t)}$. If $t_l,t_l'$ are sampled independently from $p_W$, the multi-shift estimator becomes
\begin{equation}\label{eq:shift-estimator}
\langle O\rangle_\tau
\approx
\frac{\displaystyle\sum_{l=1}^{N_s}\Phi_l\,\mu_l(O)}
{\displaystyle\sum_{l=1}^{N_s}\Phi_l\,\mu_l(I)}.
\end{equation}
Here the sample-dependent phase is
\[
\Phi_l\equiv
\frac{W(t_l)W(t_l')^*}{|W(t_l)W(t_l')|}
=e^{i[\theta(t_l)-\theta(t_l')]}.
\]
The factors $Z_W^2$ cancel in the ratio. Equation~\eqref{eq:shift-sampling} is also the normalized density used in Fig.~\ref{f1}.

A constant $f(q)$ on a bounded interval gives a sinc-like $W(t)$ with $|W(t)|\sim |t|^{-1}$, changing the long-time sampling tail from $|t|^{-2}$ to $|t|^{-3}$. The associated phase, however, winds substantially across the sampled time window. We therefore choose a smoother shift distribution whose characteristic function is
\begin{equation}\label{eq:gamma-W}
W(t)=\left(\frac{a}{a+it}\right)^k,
\end{equation}
with $a>0$ and integer $k\geq1$. Equation~\eqref{eq:gamma-W} corresponds to the normalized Gamma distribution
\[
f(q)=\frac{a^k}{\Gamma(k)}q^{k-1}e^{-aq},
\qquad q\geq0,
\]
where $\Gamma(k)$ is the Euler gamma function.
Consequently, $p_W(t)\sim |t|^{-k-2}$ at long times while all shifts are nonnegative.

The phase of Eq.~\eqref{eq:gamma-W} is
\begin{equation}\label{eq:phase}
e^{i\theta(t)}=\frac{W(t)}{|W(t)|},
\qquad
\theta(t)=-k\arctan\!\left(\frac{t}{a}\right).
\end{equation}
We scale $a$ with the natural Cauchy time scale $\tau$. The choice $a=k\tau$ gives $\theta(t)\simeq-t/\tau$ for $|t|\ll a$, so the phase variation over the dominant region $|t|\sim\tau$ remains of order unity as $\tau$ changes. At large $|t|$, $|\theta(t)|\to k\pi/2$. As a simple phase-control heuristic, requiring the asymptotic magnitude $|\theta(t)|$ to remain below $2\pi$ gives $k<4$; we therefore use the largest integer satisfying this criterion, $k=3$, which also produces a $|t|^{-5}$ sampling tail. The resulting normalized probability density and sampled phase distribution are shown in Fig.~\ref{f1}.

\begin{figure}[tbp]
\centering
\includegraphics[width=\columnwidth]{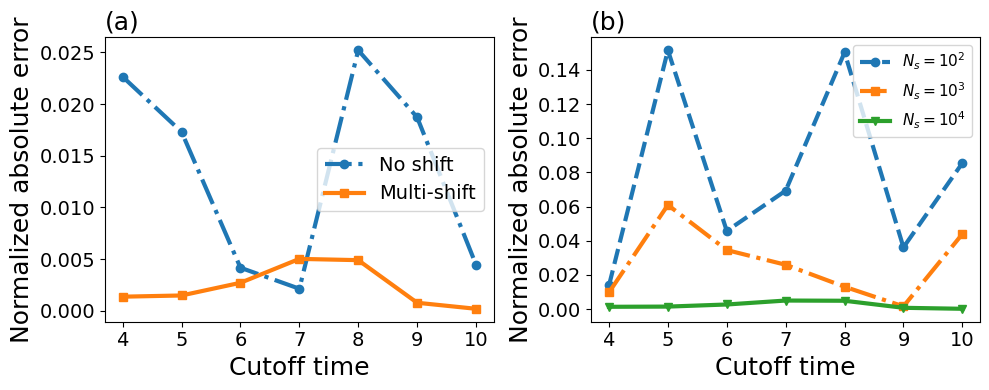}
\caption{Ground-state-energy error versus the real-time cutoff $t_m$. The results are averaged over 10 independent experiments. (a) Comparison between the original kernel and multi-shift using $N_s=10^4$ sampled time pairs. (b) Multi-shift results for $N_s=10^2$, $10^3$, and $10^4$, as indicated. The initial state is $|0\rangle^{\otimes n}=|00\cdots00\rangle$. The Ising parameters are $n=6$, $J=2$, and $h=1$, with $\tau=10$ and $W(t)$ given by Eq.~\eqref{eq:gamma-W}. The shift is chosen as $c=-E_0\simeq 11.0441$, so that the ground-state energy of $H=H_{\rm Ising}+cI$ is zero.}
\label{f2}
\end{figure}

A common application of imaginary-time evolution is ground-state preparation. In our simulations we use the open-boundary transverse-field Ising Hamiltonian
\begin{equation}\label{eq:tfim}
H_{\rm Ising}
=
-J\sum_{i=1}^{n-1}\sigma_i^x\sigma_{i+1}^x
-h\sum_{i=1}^{n}\sigma_i^z,
\end{equation}
where $n$ is the number of spins, $J$ is the nearest-neighbor Ising coupling, $h$ is the transverse field, and $\sigma_i^\alpha$ is the Pauli operator $\alpha=x,y,z$ on site $i$. For the real-time representation we use
$H=H_{\rm Ising}+cI\succeq0$ with a constant positivity shift $c$, where $\succeq0$ denotes positive semidefiniteness; the normalized ITE state is independent of $c$. For Fig.~\ref{f2}, we set $c=-E_0\simeq 11.0441$, so that the ground-state energy of the shifted Hamiltonian $H$ is exactly zero. The energy observable remains the unshifted $H_{\rm Ising}$. We quantify the ground-state-energy error by
$|E_{\rm est}-E_0|/J$, where $E_{\rm est}$ is the Monte Carlo estimate and $E_0$ is the exact ground-state energy of $H_{\rm Ising}$. The initial state is $|0\rangle^{\otimes n}=|00\cdots 00\rangle$.

As shown in Fig.~\ref{f2}, the ground-state-energy errors obtained with multi-shift are generally lower and fluctuate less across cutoff times $t_m$ than those of the original sampling kernel, consistent with the stronger suppression of long real-time evolutions in Eq.~\eqref{eq:shift-sampling}.

\section{Continuous-Variable-Assisted Multi-Shift Implementation}
\label{s4}

We next consider a fully quantum implementation assisted by one continuous-variable (CV) mode. Let $\hat x$ and $\hat p$ be canonically conjugate quadratures with $[\hat x,\hat p]=i$. Their generalized eigenstates $|x\rangle$ and $|p\rangle$ are delta normalized, and we use the Fourier convention
$\langle x|p\rangle=(2\pi)^{-1/2}e^{ipx}$. The system Hamiltonian $H$ is again taken to have nonnegative spectrum, after a constant shift if necessary. Define the normalized resource state
\[
|R(\tau)\rangle=
\sqrt{2\pi\tau}\int_{-\infty}^{\infty}dp\,
g_\tau(p)|p\rangle,
\]
where $g_\tau$ is the Cauchy density in Eq.~\eqref{eq:cauchy}. The protocol is:
\begin{enumerate}
\item \textit{Prepare the system state $|\psi_0\rangle$ (or a purification of a mixed state) and the CV resource state $|R(\tau)\rangle$.}
\item \textit{Apply the joint unitary $e^{-iH\hat p}$.}
\item \textit{Project the CV mode onto a normalized state
$|\psi_{\rm proj}\rangle=\int_{-\infty}^{\infty}dp\,\psi(p)|p\rangle$.}
\end{enumerate}
Conditioned on successful projection, the system is acted on by the Kraus operator
\[
\begin{aligned}
K_{\psi_{\rm proj}}
&=\langle\psi_{\rm proj}|e^{-iH\hat p}|R(\tau)\rangle\\
&=\sqrt{2\pi\tau}\int_{-\infty}^{\infty}dp\,
 g_\tau(p)\psi^*(p)e^{-iHp}.
\end{aligned}
\]
Hence, if $\psi^*(p)$ is proportional to a desired $W(p)$, the postselected system state realizes the corresponding MS-QITE kernel in Eq.~\eqref{eq:multishift-rt}; the proportionality constant affects the success probability but not the normalized output state.

In our previous work, a squeezed projection state was used for thermal-state preparation~\cite{Zhang2020ContinuousVariableAT}. Finite squeezing creates a tradeoff between thermal-state accuracy and postselection probability. The multi-shift viewpoint shows that one need not restrict the projection to a narrow Gaussian. Instead, one can choose a normalized projection state with broader support while preserving the desired imaginary-time map up to an overall scalar.

A particularly simple choice, with $L>0$ denoting the interval width, is the normalized step-function state
\begin{equation}\label{eq:step-state}
|\psi_L\rangle
=
\frac{1}{\sqrt L}\int_{-L}^{0}dx\,|x\rangle
=
\frac{1}{\sqrt{2\pi L}}
\int_{-\infty}^{\infty}dp\,
\frac{e^{ipL}-1}{ip}\,|p\rangle,
\end{equation}
where the $p=0$ value is understood by continuity. Its momentum wavefunction satisfies
\[
\psi_L^*(p)
=
\sqrt{\frac{L}{2\pi}}\,W_L(p),
\qquad
W_L(p)=\frac{1-e^{-ipL}}{ipL},
\]
and $W_L$ is precisely the characteristic function of the uniform positive-shift distribution
$f_L(q)=1/L$ for $0\leq q\leq L$. Substitution into the Kraus operator gives the exact ideal relation
\[
K_L
=
\frac{1-e^{-L\tau}}{\sqrt{L\tau}}\,
e^{-H\tau}.
\]
Thus $L$ changes only the postselection amplitude; finite Fock-space truncation, rather than the ideal step profile itself, is what introduces state-preparation error.

For a general initial density operator $\rho_0$, the success probability is therefore
\[
P_L
=
\frac{(1-e^{-L\tau})^2}{L\tau}\,
\operatorname{tr}\!\left(\rho_0e^{-2H\tau}\right).
\]
Maximizing the $L$-dependent prefactor is equivalent to maximizing
$(1-e^{-u})^2/u$ with $u\equiv L\tau$. The nonzero stationary point obeys
$e^u=2u+1$, giving
$u=L\tau\simeq1.25643$.

For thermal-state preparation, let
\[
\rho_\beta=
\frac{e^{-\beta H_{\rm phys}}}
{Z_\beta},
\qquad
Z_\beta\equiv\operatorname{tr}(e^{-\beta H_{\rm phys}}),
\]
where $\beta$ is the inverse temperature. Starting from the maximally mixed state
$\rho_0=I/2^n$, the normalized map
$e^{-H\tau}\rho_0e^{-H\tau}$ produces $\rho_\beta$ for $\tau=\beta/2$ when $H=H_{\rm phys}+cI$. The constant positivity shift $c$ cancels from the normalized output state, but not from the postselection probability. In particular,
\[
P_L=
\frac{(1-e^{-L\tau})^2}{L\tau}\,
\frac{e^{-\beta c}Z_\beta}{2^n},
\qquad \tau=\frac{\beta}{2}.
\]
Thus, absolute success probabilities should always be quoted together with the positivity shift used in the numerical implementation; an unnecessarily large $c$ exponentially suppresses the success rate.

For numerical implementation, both $|\psi_L\rangle$ and $|R(\tau)\rangle$ are truncated in the Fock basis. Let $m=0,1,\ldots$ denote the Fock number and define the real normalized Hermite functions
\[
\varphi_m(z)
\equiv
\frac{e^{-z^2/2}H_m(z)}
{\pi^{1/4}\sqrt{2^m m!}},
\]
where $H_m(z)$ is the physicists' Hermite polynomial. With our Fourier convention,
$\langle x|m\rangle=\varphi_m(x)$ and
$\langle p|m\rangle=(-i)^m\varphi_m(p)$. The two Fock expansions can then be written compactly as
\begin{samepage}
\begin{equation}\label{eq:fock-step}
|\psi_L\rangle
=
\sum_{m=0}^{\infty}c_m(L)|m\rangle,
\qquad
c_m(L)
=
\frac{1}{\sqrt L}\int_{-L}^{0}dx\,\varphi_m(x),
\end{equation}
\begin{equation}\label{eq:fock-resource}
\begin{aligned}
|R(\tau)\rangle&=\sum_{m=0}^{\infty}r_m(\tau)|m\rangle,\\
r_m(\tau)&=i^m\sqrt{2\pi\tau}
\int_{-\infty}^{\infty}dp\,g_\tau(p)\varphi_m(p).
\end{aligned}
\end{equation}
\end{samepage}
Equations~\eqref{eq:fock-step} and \eqref{eq:fock-resource} are equivalent to the explicit Hermite-polynomial expressions but make the normalization and basis transformation transparent. Because $g_\tau(p)$ is even, all odd Fock coefficients of the resource state vanish. In contrast, the asymmetric interval $[-L,0]$ implies that the step-function state generally contains both even and odd photon-number components. In the numerics, the sums are truncated according to a photon-number cutoff parameter $N_c$ and the resulting finite-dimensional states are renormalized.

\begin{figure}[t]
\centering
\includegraphics[width=\columnwidth]{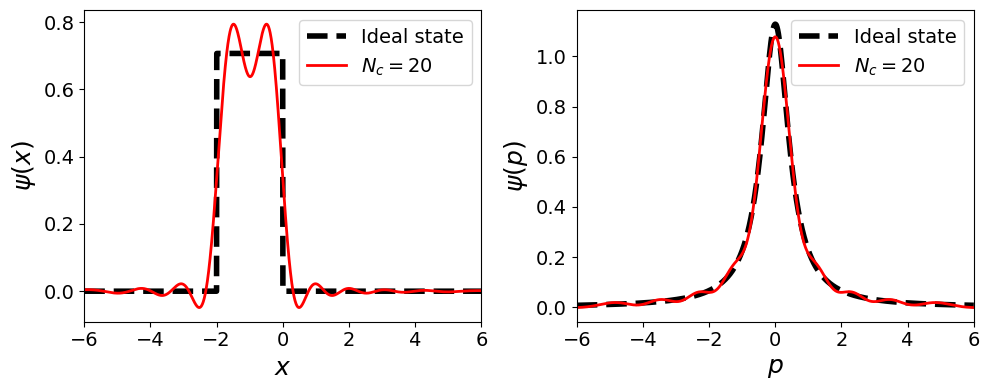}
\caption{Finite-Fock-space approximation of the projection and resource states. Left: position-space wavefunction of the step-function state $|\psi_L\rangle$. Right: momentum-space wavefunction of $|R(\tau)\rangle$. Dashed curves denote the ideal states and solid curves the truncated states for $N_c=20$.}
\label{f3}
\end{figure}

With a finite Fock cutoff, these states can be synthesized in a cavity as finite superpositions. Arbitrary Fock-state superpositions can be generated using sequences of qubit rotations and Jaynes--Cummings-type qubit--cavity couplings~\cite{Law1996ArbitraryCO,Hofheinz2009SynthesizingAQ}, or using number-dependent arbitrary phase gates together with displacement operations in the dispersive regime~\cite{Heeres2015CavitySM,PhysRevLett.118.223604}. Representative finite-cutoff approximations are shown in Fig.~\ref{f3}.

\begin{figure}[t]
\centering
\includegraphics[width=\columnwidth]{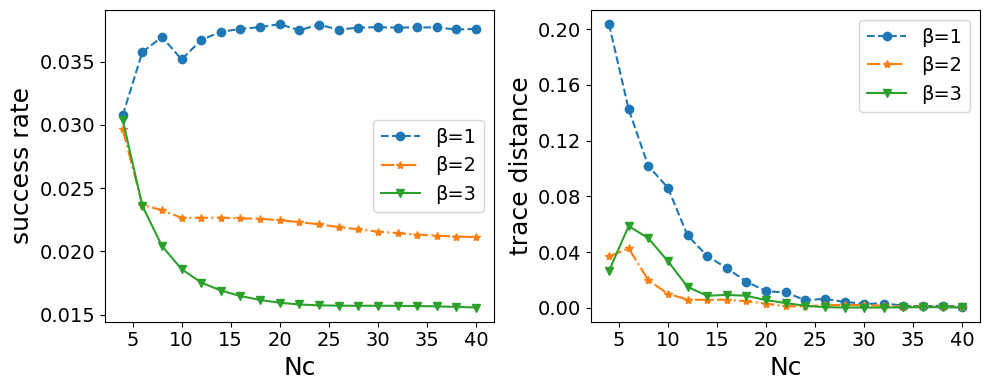}
\caption{Convergence with the Fock-space cutoff $N_c$ for thermal-state preparation using $|\psi_L\rangle$ as the projection state. Left: postselection success probability. Right: trace distance from the exact Gibbs state. Curves are shown for $\beta=1$, $2$, and $3$. The Ising parameters are $n=4$, $J=1$, and $h=1$. In the CV evolution we use $H=H_{\rm Ising}+5I$, which makes the ground-state energy slightly positive. The step-state parameter is chosen at the theoretical optimum, $L\tau=1.25643$.}
\label{f4}
\end{figure}

\begin{figure}[tbp]
\centering
\includegraphics[width=\columnwidth]{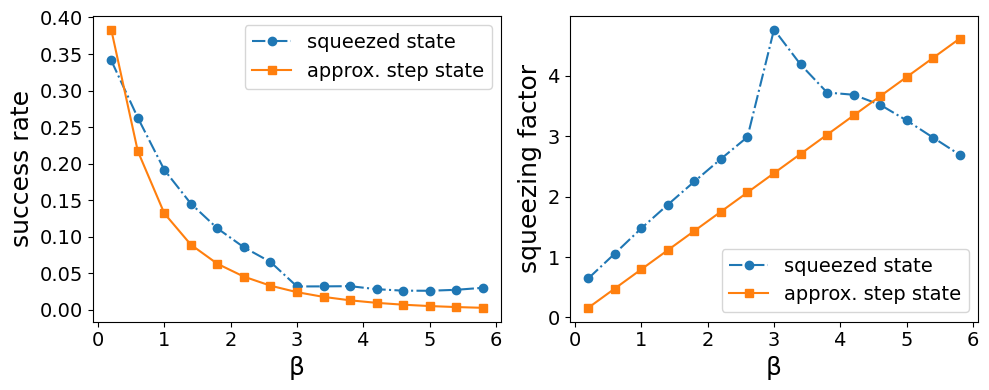}
\caption{Resource comparison at matched thermal-state accuracy. Left: postselection success probability. Right: squeezing factor. The orange curves use the displaced squeezed approximation to the step-function state with $s_{\rm step}=2/L$ and the optimal $L\tau=1.25643$; for each $\beta$, the blue squeezed-state baseline is tuned to the same trace distance. The Ising parameters are $n=4$, $J=1$, and $h=1$, and the CV evolution uses the same shift $H=H_{\rm Ising}+5I$ as in Fig.~\ref{f4}.}
\label{f5}
\end{figure}

We next study thermal-state preparation for a four-qubit transverse-field Ising model. We quantify the state error by the trace distance
$D_{\rm tr}(\rho,\sigma)=\frac12\|\rho-\sigma\|_1$, where $\|A\|_1\equiv\operatorname{tr}\sqrt{A^\dagger A}$ is the trace norm. For the data in Figs.~\ref{f4} and~\ref{f5}, the Hamiltonian used in the CV evolution is shifted as $H=H_{\rm Ising}+5I$. For $n=4$ and $J=h=1$, the unshifted ground-state energy is $E_0\simeq-4.7588$, so the shifted ground-state energy is $E_0+5\simeq0.2412>0$, while the normalized Gibbs state is unchanged.

As shown in Fig.~\ref{f4}, the success probability approaches a stable value as $N_c$ increases, while the trace distance converges toward zero. To compare with the squeezed-state method of Ref.~\cite{Zhang2020ContinuousVariableAT}, we use the convention $\langle p|0,s\rangle\propto s^{-1/2}e^{-p^2/(2s^2)}$, for which larger $s$ means stronger squeezing in $x$. We approximate the ideal step profile by a displaced squeezed state centered at $x=-L/2$ and choose $s_{\rm step}=2/L$, so that its characteristic position-space width is comparable to the interval length $L$. At the optimal $u_\star=L\tau=1.25643$ and with $\tau=\beta/2$, this gives
\[
s_{\rm step}=\frac{\beta}{u_\star}\simeq0.796\,\beta,
\]
which explains the linear orange curve in the right panel of Fig.~\ref{f5}. At each $\beta$, the squeezing factor of the original squeezed-state protocol is then adjusted until its output thermal state has the same trace distance as the approximate-step construction. The two success probabilities and the squeezing factors can therefore be compared at matched state accuracy.

As shown in Fig.~\ref{f5}, the approximate step-function construction requires less squeezing over an intermediate range of inverse temperatures. This illustrates how the multi-shift freedom can be used to trade the shape of the projection state against implementation resources without changing the target normalized imaginary-time evolution.

\section{Conclusion}
\label{s5}

We have introduced multi-shift quantum imaginary-time evolution (MS-QITE) as an optimization framework for quantum implementations of imaginary-time evolution. The key observation is that a distribution of energy shifts changes only an overall scalar in imaginary time but modifies the interference kernel in a real-time representation. In the Monte Carlo realization, this freedom suppresses long-time samples while keeping the phase fluctuations controlled. In the continuous-variable-assisted realization, it maps directly to the choice of projection wavefunction and permits finite-support states that can improve the squeezing--success-probability tradeoff. Numerical simulations for transverse-field Ising models illustrate both mechanisms. The framework suggests several natural extensions, including optimization of shift distributions under hardware-specific cost functions, systematic accounting of state-preparation overhead, and scaling tests in larger systems.

\begin{acknowledgments}
This work was supported by the National Natural Science Foundation of China (Grant Nos.~12375013 and 12547109) and by the Guangdong Provincial Quantum Science Strategic Initiative (Grant No.~GDZX2503008).   
\end{acknowledgments}

\bibliography{Mybib_revised_submission_v2}
\end{document}